\documentclass[%
 preprint,
 amsmath,amssymb,
aps,
prb,
superscriptaddress
]{revtex4-1}
\usepackage{graphicx}
\usepackage{dcolumn}
\usepackage{bm}
\usepackage{here}
\usepackage{braket}
\usepackage{url}
\usepackage{comment}

\usepackage{color}

\usepackage{multirow}

\begin{document}

\title{High-Throughput Mapping of Magnetic Properties via the on-the-fly XMCD spectroscopy in a Combinatorial Fe-Co-Ni Film}

\author{Y. Yamasaki} 
\affiliation{Center for Basic Research on Materials (CBRM), National Institute for Materials Science (NIMS), Tsukuba, Ibaraki, 305-0047, Japan}
\affiliation{International Center for Synchrotron Radiation Innovation Smart, Tohoku University, Sendai 980-8577, Japan}
\affiliation{Center for Emergent Matter Science (CEMS), RIKEN, Wako 351-0198, Japan}

\author{N. Sasabe} 
\affiliation{Center for Basic Research on Materials (CBRM), National Institute for Materials Science (NIMS), Tsukuba, Ibaraki, 305-0047, Japan}

\author{Y. Ishii} 
\affiliation{Center for Basic Research on Materials (CBRM), National Institute for Materials Science (NIMS), Tsukuba, Ibaraki, 305-0047, Japan}

\author{Y. Sekiguchi}
\affiliation{Center for Basic Research on Materials (CBRM), National Institute for Materials Science (NIMS), Tsukuba, Ibaraki, 305-0047, Japan}

\author{A. Sumiyoshiya} 
\affiliation{General Incorporated Foundation Photon Science Innovation Center (PhoSIC), 468-1 Aramaki-aza Aobayama Universe 306, Aoba-ku, Sendai-shi, Miyagi 980-0845, Japan}

\author{Y. Tanimoto} 
\affiliation{General Incorporated Foundation Photon Science Innovation Center (PhoSIC), 468-1 Aramaki-aza Aobayama Universe 306, Aoba-ku, Sendai-shi, Miyagi 980-0845, Japan}

\author{Y. Kotani} 
\affiliation{General Incorporated Foundation Photon Science Innovation Center (PhoSIC), 468-1 Aramaki-aza Aobayama Universe 306, Aoba-ku, Sendai-shi, Miyagi 980-0845, Japan}

\author{T. Nakamura} 
\affiliation{International Center for Synchrotron Radiation Innovation Smart, Tohoku University, Sendai 980-8577, Japan}

\author{H. Nomura} 
\affiliation{Center for Basic Research on Materials (CBRM), National Institute for Materials Science (NIMS), Tsukuba, Ibaraki, 305-0047, Japan}
\affiliation{International Center for Synchrotron Radiation Innovation Smart, Tohoku University, Sendai 980-8577, Japan}

\begin{abstract}
High-throughput X-ray magnetic circular dichroism (XMCD) spectroscopy was conducted on Fe-Co-Ni compositionally graded films to systematically analyze the variation of magnetic properties as a function of composition. 
On-the-fly XMCD measurement enabled rapid spectral acquisition.
Measurement time was reduced approximately tenfold compared to conventional stepwise methods, while maintaining high precision. 
The obtained XMCD spectra were processed using the Savitzky-Golay denoising technique, and element-specific magnetic properties were extracted using XMCD sum rules. 
By mapping the orbital and spin magnetic moments across the composition gradient, we identified key regions exhibiting enhanced soft magnetic properties. 
This study demonstrates the effectiveness of high-throughput synchrotron-based spectroscopy for accelerating materials discovery and optimizing functional magnetic materials.
\end{abstract}

\maketitle

\section{Introduction}
Magnetic materials exhibit various properties depending on their composition, electronic state, and lattice structure. 
Understanding these properties is crucial for optimizing material performance in magnetic storage, spintronics, and next-generation magnetic devices. 
Among various magnetic materials, Fe-Co-Ni alloys have attracted significant attention due to their tunable magnetic anisotropy, high saturation magnetization, and excellent mechanical properties \cite{Chesnutt1993-vy,Osaka1998-bz}. 
These make them promising candidates for spintronic devices and magnetic recording media.
One of the key characteristics of Fe-Co-Ni alloys is their composition-dependent magnetic properties. According to the Slater-Pauling rule, Fe-Co alloys exhibit the highest saturation magnetization, which peaks at approximately Fe$_{65}$Co$_{35}$, reaching a value close to 2.4 $\mu_B$/atom—higher than that of pure Fe (2.2 $\mu_B$/atom) or Co (1.7 $\mu_B$/atom) \cite{Iwasaki2021-yl}. 
Moreover, by carefully adjusting the composition, soft magnetic materials with near-zero magnetostriction have been achieved, which is beneficial for minimizing energy losses in magnetic applications \cite{Chesnutt1993-vy, Osaka1998-bz}. 

The lattice structure of Fe-Co-Ni alloys plays a crucial role in determining their electronic and magnetic properties. Each element exhibits a distinct lattice structure: Ni adopts a face-centered cubic (FCC, space group $Fm\bar{3}m$) structure, Fe has a body-centered cubic (BCC, $Im\bar{3}m$) structure, and Co crystallizes in a hexagonal close-packed (HCP, $P6_3/mmc$) structure. The crystal structure varies with the Fe-Co-Ni composition ratio, leading to phase transformations that significantly impact the material’s physical properties. 
The ternary phase diagram of lattice structure in Fe-Co-Ni alloys has been determined as illustrated in Figure~\ref{fig1}(a) \cite{Yoo2006-oa}. 
These structural differences influence electronic interactions and magnetic behaviors, including exchange coupling, magnetocrystalline anisotropy, and spin-orbit interactions.

Recent advancements in combinatorial materials science have enabled systematic investigations of composition-dependent properties within a single sample \cite{Green2013-sd, Tan2024-wc}. 
In combinatorial samples, high-throughput methods such as magneto-optical Kerr effect (MOKE) measurements and micro X-ray diffraction (XRD) have been utilized to rapidly determine the magnetic properties and crystal structures for various compositions \cite{Iwasaki2017-dq}. 
However, to gain deeper insight into the origin of these magnetic properties \cite{Soderlind1992-jp}, spectral measurements using synchrotron radiation are essential in addition to laboratory-based experiments. Such techniques enable the precise evaluation of element-specific electronic and chemical states, which are critical for understanding the fundamental mechanisms governing the material’s behavior.

In addition, X-ray magnetic circular dichroism (XMCD) measurements provide valuable insights into the local magnetic properties \cite{Nishio2022-ml}. 
The XMCD sum rules, the integration of the XMCD spectrum, enable separate measurement of spin and orbital magnetic moments. 
Orbital angular momentum, through spin-orbit coupling, determines magnetic anisotropy and provides valuable insights into the origins of soft magnetic properties. Moreover, recent studies have shown that XMCD can also detect the magnetic dipole term, which is responsible for the anomalous Hall effect observed in antiferromagnets \cite{Yamasaki2020-lr, Hayami2021-xf, Kimata2021-bd}.
These techniques enable exploration of material phase diagrams and optimization of their magnetic properties for practical applications.
In this study, we focus on the rapid spectral measurement of Fe-Co-Ni compositionally graded films using the on-the-fly XMCD spectrum measurement. 
By employing a high-throughput measurement approach, we aim to elucidate the correlation between composition and electronic/magnetic properties contributing to the development of next-generation spintronic and magnetic storage materials.

\section{Experiments}
\begin{figure*}[t]
    \centering
    \includegraphics[width=0.95\textwidth]{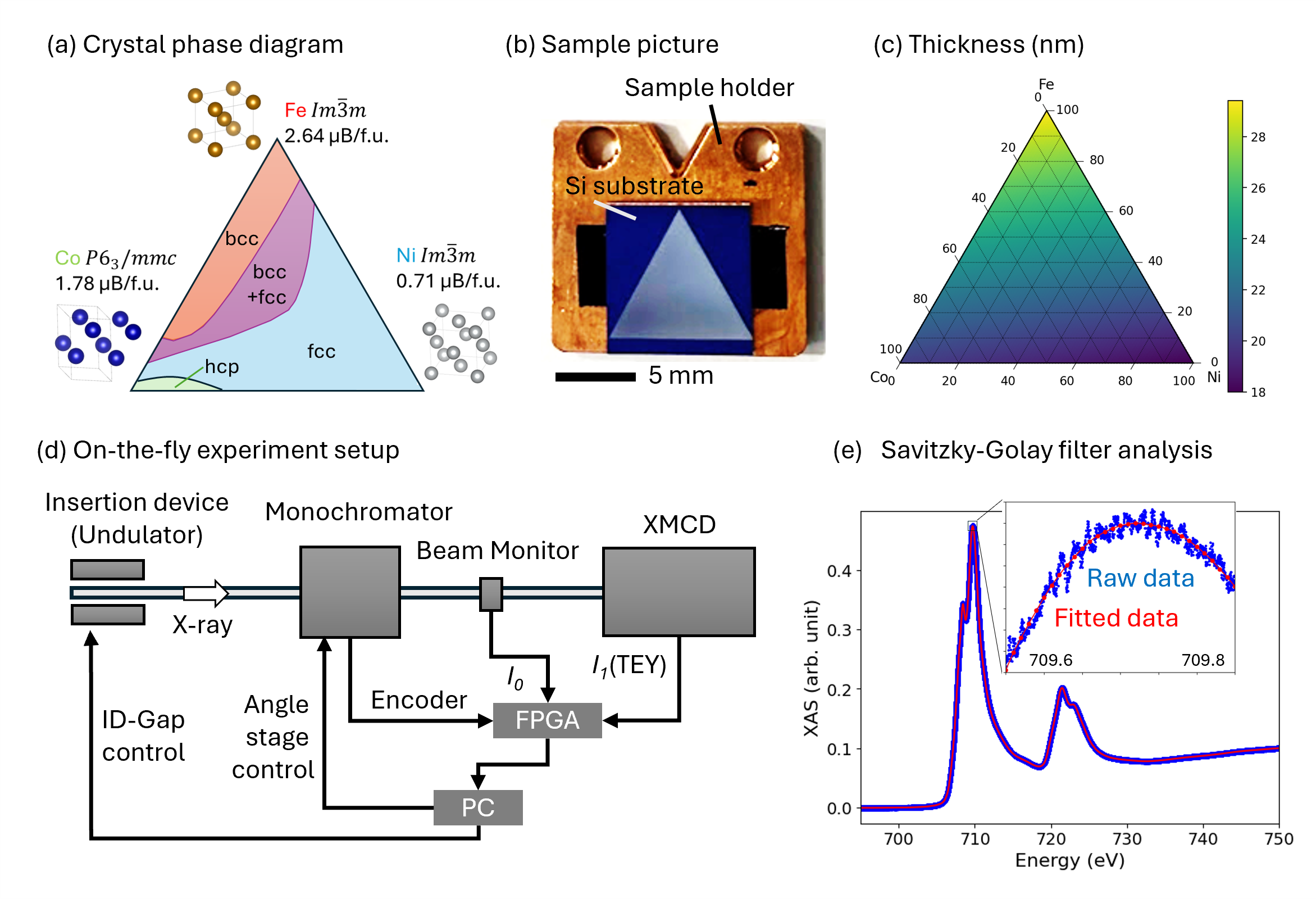}
    \caption{(a) Ternary lattice phase diagram for Fe-Co-Ni alloy. 'bcc', 'fcc', and 'hcp' represent the body-centered cubic, face-centered cubic, and hexagonal close-packed structure, respectively. (b) Picture of a combinatorial Fe-Co-Ni thin film fabricated by the sputtering method. The scale bar is equivalent to 5 mm. (c) Thickness of fabricated sample. (d) Experimental configuration of the on-the-fly XMCD experiment. (e) An example of the Savitzky-Golay filter denoising analysis for XAS around Fe $L$-edge absorption.}
    \label{fig1}
\end{figure*}

Fe$_x$Co$_y$Ni$_z$ ($x+y+z=1$) composition gradient film, a combinatorial film, was fabricated using the Combinatorial Magnetron Sputtering System (CMS-6410) by COMET Inc., Tsukuba, Japan.
The sample was designed in a triangular shape with a side length of 7.28 mm, allowing for a continuous composition gradient of the three elements, $i.e.$ Fe, Ni, and Co [Figure \ref{fig1}(b)]. 
For these metal targets, 3N purity materials were used to ensure high-quality film deposition.
The deposition rates of each element were adjusted so that their concentrations were equal at the center of the sample.
The film was deposited on a thermally oxidized Si substrate to ensure a stable and uniform growth surface.
The deposition process was carried out at 400 $^\circ$C under a vacuum pressure of \(2.7 \times 10^{-5}\) Pa.
The final thicknesses of the films at the Fe, Co, and Ni corners were 29.4 nm, 20.1 nm, and 18 nm, respectively.
Figure \ref{fig1}(c) shows the thickness of each composition interpolated linearly from the thicknesses at the sample corners.

X-ray absorption spectroscopy (XAS) measurements were performed using the BL14U beamline at NanoTerasu, Sendai, Japan.
The energy range of the XAS measurements was set from 650 eV to 950 eV to cover the $L_3$ and $L_2$ absorption edges of Fe, Co, and Ni.
The XAS measurements were conducted using the Total-Electron Yield (TEY) method, which provides surface-sensitive detection of the XMCD signal by collecting emitted electrons.
The spatial dependence of XAS was measured using the Scanning X-ray Microscopy (SXM) system installed at the beamline \cite{Kotani2018-ib}.
The sample position was measured in 0.5 mm steps controlled by the linear stage stepping motors, and to cover the entire triangular area, XAS measurements were performed at 66 different locations.
The beam size at the sample position was approximately 0.5 mm × 0.5 mm.

The conventional XAS measurements are typically performed by stopping the monochromator at each energy point and measuring the signal before moving to the next energy.
While this stepwise approach provides high accuracy, it significantly increases measurement time, making high-throughput measurements preferable for systematic studies of composition-dependent properties.
The present study measured the energy spectrum by continuously varying the monochromator while collecting signals on-the-fly, acquiring approximately 100k sampling points per second using the Field-Programmable Gate Array (FPGA), as shown in the experimental setup diagram in Fig. \ref{fig1}(d).
After the control PC sends the energy scan command, the TEY signal ($I_1$) and the incident X-ray beam monitor signal for normalization ($I_0$) are each passed through high-speed current amplifiers and sampled at high speed along with the encoder signal of the monochromator angle that corresponds to the energy of the incident x-ray.
At this beamline, soft X-ray photons are generated by a twin-helical undulator.
During the on-the-fly measurement, the gap of the magnetic arrays of the undulator corresponding to the selected photon helicity was also dynamically controlled with the grating pitch scan to synchronize with the photon energy.

This method increases measurement speed but also introduces more noise than conventional stepwise measurements due to the continuous motion of the optical elements and short signal acquisition time window [Fig. \ref{fig1}(e)].
This issue will be addressed by performing the denoising filter processing analysis discussed later.
This approach allows for efficient data acquisition and significantly reduces the total measurement time while maintaining high spectral resolution.
On-the-fly XAS measurements were performed for left- and right-circularly polarized light, corresponding to the spin angular momentum of light $p=\pm \hbar$.
To measure the saturated magnetization state, a magnetic field of 1 T was applied using a superconducting magnet mounted on the SXM in the same direction as the incident X-rays \cite{Kotani2018-ib}.
To eliminate artifacts, one XMCD spectrum was derived by analyzing four XAS spectra obtained under different polarizations ($p=\pm \hbar$) and magnetic fields ($h=\pm 1$ T) combinations.
Consequently, for each of the 66 sample positions, measurements were performed under different magnetic field and polarization conditions, and a total of 264 XAS spectra covering the absorption edges of Fe, Co, and Ni were acquired within 16 hours.

\section{Analysis}
The obtained XAS spectra were processed using the Savitzky-Golay filter to remove noise while preserving spectral features. 
This filter applies a least-squares polynomial fit over a rolling window of data points, effectively smoothing the data while maintaining spectral details.
Unlike simple moving averages, the Savitzky-Golay filter preserves peak structures and high-frequency features, making it ideal for spectral analysis in XAS measurements. 
It functions as a low-pass filter, reducing noise while maintaining sharp transitions in the spectrum.

The core idea of the Savitzky-Golay filter is to approximate the data points within a moving window of aperture size $w$ by a polynomial of a certain degree $p$. 
The value of the polynomial at the central point of the window is then taken as the smoothed value. 
This process is repeated for each point in the dataset, resulting in a smoothed signal.
Suppose we have a set of data points $(x_i, y_i)$ where $i$ ranges from 1 to $w$ and it is centered at $x_k$. 
These observed data points are aimed to fit by a polynomial of degree $p$, which can be expressed as:
\begin{equation}
    y = f_{\bm{\alpha}}(x) = \alpha_0 + \alpha_1 x + \alpha_2 x^2 + \dots + \alpha_p x^p.
\end{equation}
For the given window of data points, we determine the coefficients $\bm{\alpha}=(\alpha_0, \alpha_1, \dots, \alpha_p)$ such that the polynomial best fits the data points. 
This is achieved by minimizing the sum of the squares of the differences between the actual data points \(y_i\) and the polynomial values:
\begin{equation}
    \hat{\bm{\alpha}} = \underset{{\bm{\alpha}}} {\operatorname{argmin}} \sum_{i=1}^{w} \left( y_{i} - \sum_{j=0}^{p} \alpha_j x_{i}^j \right)^2.
\end{equation}
Then, the filtered value is obtained as $\hat{y}_k = f_{\hat{\bm{\alpha}}}(x_k)$.
This least-squares minimization ensures that the polynomial smoothly follows the data trend while preserving sharp transitions.

Figure \ref{fig1}(e) shows the raw data of normalized XAS spectra ($I_1/I_0$) with the incident X-ray intensity $I_0$ and the TEY signal of the sample $I_1$ (the blue scattered points) at around Fe $L$-edge energy region.
Due to the short measurement time aperture, the data contain significant noise. 
Additionally, since the monochromator is continuously moving, long-period oscillatory artifacts appear due to its feedback controller. 
By applying curve fitting using the Savitzky-Golay filter to datasets, we obtained denoised data as shown in the red line of Fig. \ref{fig1}(e).
In this study, the Savitzky-Golay filter was implemented using the SciPy library's \texttt{savgol\_filter} function to process raw XAS data.
The Savitzky-Golay filter analysis was applied with the fitting parameters of $p = 2$ and $w = 1025$,  corresponding to a curve fit over an approximate energy range of 0.15 eV.
After performing noise reduction analysis, the XMCD calculation was carried out using the four XAS spectra data sets ($p=\pm \hbar$ and $h =\pm 1$ T).

\section{Results}
\begin{figure*}[t]
    \centering
    \includegraphics[width=0.95\textwidth]{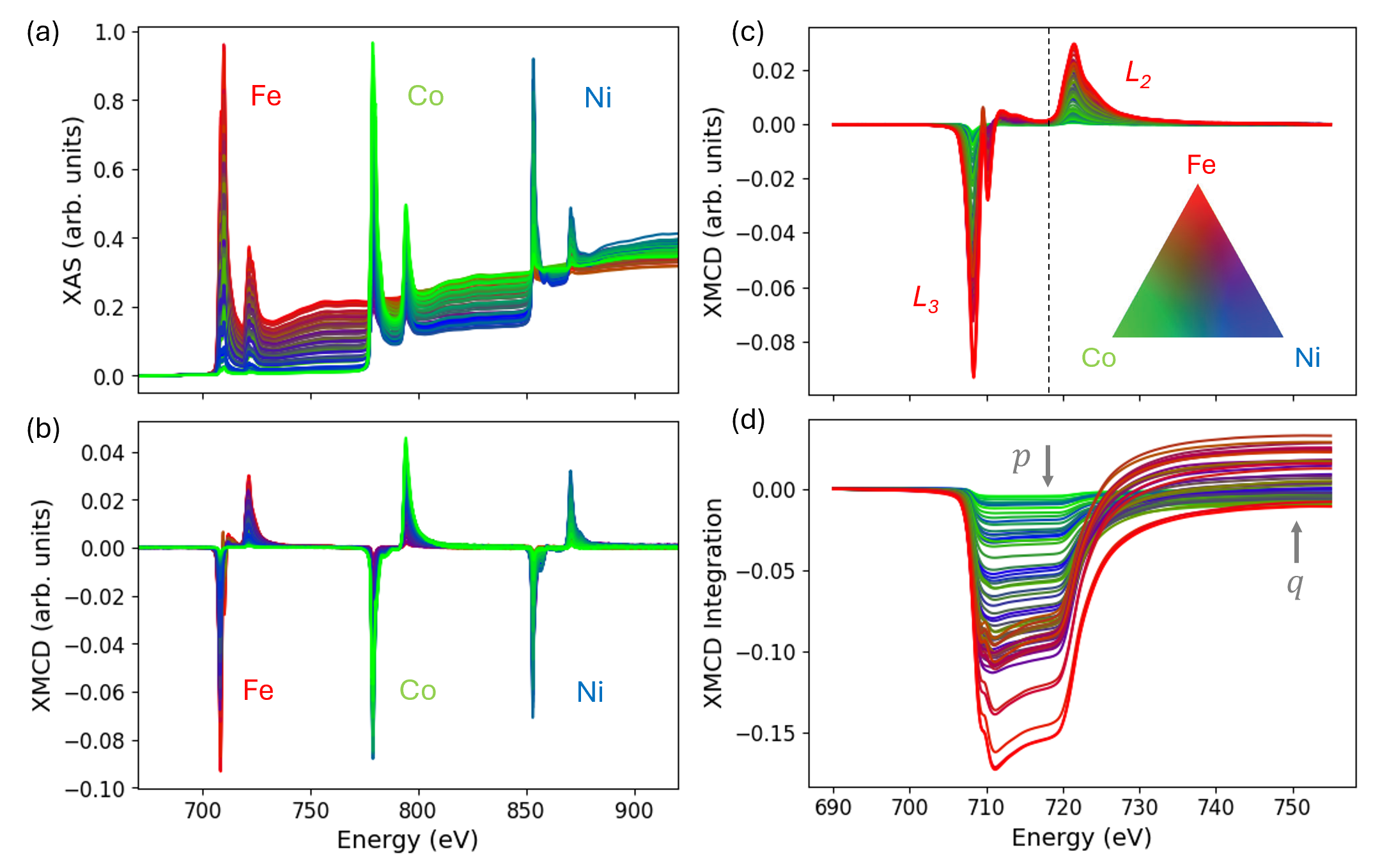}
    \caption{X-ray absorption spectrum (XAS) and x-ray magnetic circular dichroism (XMCD) for Fe in the Fe–Ni–Co combinatorial film. (a) Normalized TEY-XAS and XMCD spectra covering the Fe $L_3$ and $L_2$ absorption edges at multiple sample positions. Each line is colored according to its ternary (Fe–Ni–Co) composition via the RGB mapping inset in (b). (b) Zoomed view of the XMCD at Fe $L$-edge and its integrals of the XMCD signal. $p$ and $q$ indicate the integration value at 718 eV and 750 eV, respectively. These integrals feed into the sum-rule formulas for extracting the orbital and spinful magnetic dipole moments.}
    \label{fig:xas_xmcd}
\end{figure*}

Figure \ref{fig:xas_xmcd}(a) and \ref{fig:xas_xmcd}(b) respectively show XAS and XMCD spectra covering $L_3$ and $L_2$ absorption edges for Fe, Co, and Ni measured with the changing position of the combinatorial sample.
Figure 2(c) presents an enlarged view of the Fe XMCD region, while Figure 2(d) shows the corresponding integrated values.
The line colors are determined using the RGB color function based on the ternary coordinates, corresponding to the positions on the colormap shown in the inset of Fig. \ref{fig:xas_xmcd}(c).
The composition varies depending on the position of the combinatorial sample, resulting in changes in the XAS and XMCD spectra. This enables element-specific evaluation of magnetic properties for each composition.
To correlate each spectrum with physical properties, analyses based on the sum rules of X-ray absorption are performed. 
X-ray absorption can be derived from Fermi’s golden rule for electric dipole transitions, reflecting the transition probabilities depending on the magnetic quantum numbers of the valence states. 
Therefore, by calculating spectra for multiple polarizations, it becomes possible to extract magnetic properties \cite{Thole1992-ne,Carra1993-bg}.
Notably, the magnetic and electric multipoles accessible via the sum rules depend on the chosen polarization.
Specifically, XAS sum rules probe electric monopoles, XMCD probes magnetic dipoles, and XMLD probes electric quadrupoles, with both spinless and spinful multipoles measurable selectively in each case.
 \cite{Laan1998-hk, Kusunose2020-od, Hayami2021-xf, Yamasaki2020-lr}.

\begin{figure*}[t]
    \centering
    \includegraphics[width=0.98\textwidth]{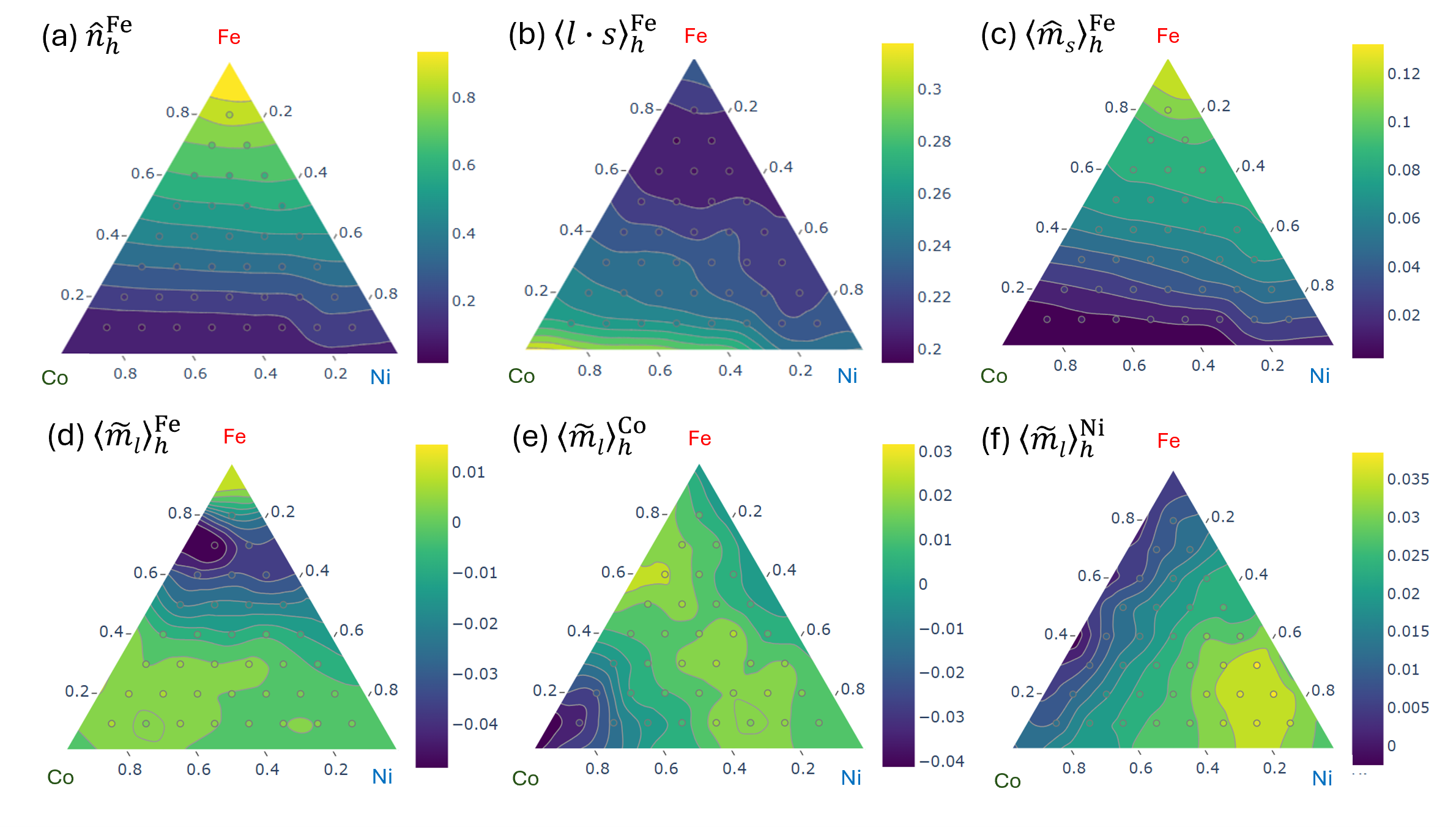}
    \caption{Element‑specific compositional and magnetic‑moment maps of the Fe–Ni–Co gradient film. 
    (a) The hole number of Fe obtained from integration TEY‑XAS, plotted on a ternary diagram (66 measurement points with the use of the cartesian-space interpolation).
    (b) Map of the spin–orbital inner product $\bm{l}\cdot\bm{s}$ for Fe obtained by the XAS sum rule, which gauges the effective spin–orbit coupling. 
    (c) The spinful magnetic dipole terms $\langle m_s^\text{Fe}\rangle$ evaluated by the XMCD sum rule.
    (d), (e), and (f) show orbital magnetic moment $\langle \tilde{m}\rangle_h^A$ of $A=$ Fe, Co, and Ni, respectively.}
    \label{fig:ternarymap}
\end{figure*}

The summation of XAS for linear ($||z$) and circular polarizations ($\mu_\pm$) provides information on the electronic monopole, \textit{i.e.} the hole concentration $n_h$ and the inner product of orbital and spin angular momenta $\bm{l}\cdot{s}$. 
Assuming that the sample is isotropic, $i.e.$ the crystal anisotropy and substrate strain are minimal, the following equation can be used as an approximation: $S_{00} \equiv \mu_++\mu_-+\mu_z \sim \frac{3}{2}(\mu_++\mu_-)$.
The number of holes $n_h^A$ ($A=\text{Fe, Co, Ni}$) can be determined from the total X-ray absorption integration at $L_3$ and $L_2$ edges:
\begin{equation}\label{hole_sum}
 \langle n_h^A\rangle= C\int_{A,L_3+L_2} S_{00} dE,
\end{equation}
where $C$ is a scaling factor depending on the experimental condition.
Analyzing the XAS integration, we can obtain the normalized hole number in the valence state of Fe, $\langle\hat{n}_h^\text{Fe}\rangle\equiv\langle n_h^\text{Fe}\rangle/\langle n_h^\text{total}\rangle$ with $n_h^\text{total}\equiv n_h^\text{Fe}+n_h^\text{Co}+n_h^\text{Ni}$, as shown in the ternary plot of Fig. \ref{fig:ternarymap}(a).
It can be seen that the hole occupancy ratio of Fe varies according to the nominal Fe composition.

On the other hand, the spinful electric monopole, the inner product of orbital $\bm{l}$ and spin angular momentum $\bm{s}$, can be determined from the difference X-ray absorption integral between $L_3$ and $L_2$ edges:
\begin{equation}
\langle \bm{l}^A\cdot\bm{s}^A\rangle= C\left(\int_{A,L_3} S_{00} dE-2\int_{A,L_2}S_{00}dE\right),
\end{equation}
where $\bm{l}^A$ and $\bm{s}^A$ represent orbital and spin angular momenta on the valence state of element $A$.
Figure \ref{fig:ternarymap} (b) shows the spinful electronic monopole normalized by the hole number, defined as $\langle \bm{l}\cdot\bm{s}\rangle_{h}^\text{Fe}\equiv \langle \bm{l}^\text{Fe}\cdot\bm{s}^\text{Fe}\rangle/\langle n_h^\text{Fe} \rangle$.

XMCD sum rules allow for the quantitative determination of the orbital and spin magnetic moments. 
These sum rules, formulated by Thole \textit{et al.} \cite{Thole1992-ne} and Carra \textit{et al.} \cite{Carra1993-bg}, provide a method to extract magnetic moments from XMCD spectra at the $L_{2,3}$ edges.
The spinless sum rule gives information about orbital angular momentum, which is given by
\begin{equation}
\langle {l}_z^A\rangle= -2C\int_{A,L_3+L_2} S_{10} dE,
\end{equation}
with $S_{10}\equiv \mu_+ - \mu_-$ and $C$ being the same coefficient as Equation \ref{hole_sum}.
In contrast, the spinful sum rule provides the information of spin ${s}_z^A$ and the anisotropic magnetic dipole term ${t}_z^A$, expressed as
\begin{equation}
\langle m_z^A\rangle= -3C\left(\int_{A,L_3} S_{10} dE-2\int_{A,L_2}S_{10}dE\right),
\end{equation}
with $\langle m_z^A\rangle =2\langle {s}_z^A\rangle + 7\langle {t}_z^A\rangle$.
The $t_z$ term, defined as the product of the electric quadrupole operator and the spin operator \cite{Oguchi2004-ms}, becomes the dominant contribution to XMCD in antiferromagnets that exhibit magnetic dipole symmetry \cite{Yamasaki2020-lr,Sasabe2021-vz,Kimata2021-bd,Sasabe2023-gf,Yamasaki2025-yd}. 
In strongly anisotropic regions, such as the sample surface, its magnitude increases markedly, so in surface‑sensitive TEY measurements the $t_z$ contribution cannot be ignored, and extracting a purely spin‑derived quantity is therefore difficult.

In the Fig. \ref{fig:xas_xmcd}(d), $p$ and $q$ represent the integrated values at 718 eV, which is the boundary between the $L_3$ and $L_2$ edges, and at 750 eV, which is located sufficiently above the $L_3$ edge, respectively.
Using these values, the orbital magnetic moment and the spinful magnetic dipole are given by $\langle l_z^\text{Fe} \rangle = - 2Cq$ and $\langle m_z^\text{Fe}\rangle = -3(3p-2q)C$, respectively \cite{Chen1995-nl,Nakamura2014-rx}.
Figure \ref{fig:ternarymap}(c) shows the spinful magnetic dipole normalized by the total hole number, defined by $\langle \hat{m}_s\rangle_h^\text{Fe}\equiv \langle m_s^\text{Fe}\rangle/\langle n_h^\text{total}\rangle$.
For orbital angular momenta, we show the ternary plots of the ratio between the orbital ($l_z$) and magnetic dipole ($s_z$ and $t_z$) terms, defined by 
$\langle \tilde{m}_l\rangle_{h}^A\equiv 
\frac{\langle l_z^A\rangle}{\langle m_s^A\rangle}\frac{\langle n_h^A \rangle}{\langle\hat{n}_h^\text{total}\rangle}$, for $A=$Fe, Co, and Ni in Figs \ref{fig:ternarymap}(d), (e), and (f), respectively.
Here, since in regions of low elemental concentration ${\langle m_s^A\rangle}$ becomes weak, resulting in increased noise in the data, we multiply by the normalized hole count to plot the essential information of the orbital angular momentum.

\section*{Discussion}
Element‐specific mapping by XAS and XMCD makes it possible to visualize the nominal composition of each element across the ternary library. 
As displayed in Fig. \ref{fig:ternarymap}(a), the Fe distribution broadly follows the intended composition rate, yet a slight deviation appears in the Ni‑rich corner. 
Since total‑electron‑yield (TEY) XAS probes only a few nanometres from the surface, we tentatively attribute this mismatch to Fe ions that have segregated at the film surface; however, the precise origin of the anomaly remains unresolved.
The inner product of spin and orbital moments, which reflects the strength of the Fe spin–orbit interaction, decreases markedly in the Fe‑(20–40 at \%)Co/Ni region, as shown in Fig. \ref{fig:ternarymap}(b), consistent with the soft‑magnetic behavior reported for Permalloy‑type alloys. 
Conversely, the signal increases around the Co‑rich sector, mirroring the enhancement expected for hcp Co, where lower symmetry amplifies spin–orbit coupling. 
Ni exhibits an even wider region of reduced interaction, again compatible with the low magnetic anisotropy of Fe–Ni soft magnets.

For Fe, the spinful magnetic dipole terms $\langle m_s\rangle$ trace the nominal composition overall, yet its magnitude is noticeably larger in Fe–Ni alloys than in Fe–Co ones. 
By contrast, the orbital magnetic moment, although small in the pure Fe, grows when Co and/or Ni are introduced. 
In that composition window, earlier reports attribute the high total moment to an increased spin and orbital contribution, but our data suggest that the rise seems to originate only from the orbital component. 
Since the spinful magnetic dipole term in XMCD sum rule is not a pure spin observable \cite{Wu1994-ow}, its contribution may fail to capture the true moment enhancement. 
A rigorous separation of spin ($s_z$) and anisotropic magnetic dipole ($t_z$) parts, $e.g.$ by angular‑dependent XMCD or comparison with first‑principles calculations, is therefore essential for an unambiguous interpretation.
For Co and Ni, the orbital moment scales almost monotonically with their respective concentrations, as shown in Figs. \ref{fig:ternarymap}(e) and \ref{fig:ternarymap}(f). 
Fe and Co share an extended region of suppressed orbital angular momentum, faithfully reflecting the soft‑magnetic character of this ternary alloy family.

Since the XMCD signal was acquired in total-electron-yield (TEY) mode, which probes only the top few nanometres, it should be noted that the measurement is intrinsically sensitive to surface chemistry.
Indeed, the results of Fe $L_3$-edge line shapes depart from the transmission-mode spectrum of iron \cite{Chen1995-nl}, displaying additional features characteristic of iron oxides.
These observations indicate that a thin, unintentionally oxidized surface layer and possible deviations from the designed composition contaminate the TEY signal, so quantitative moment determinations must be interpreted cautiously.
The anomalous $L_3$-edge profile due to the surface oxidation also causes the counter-intuitive sum-rule results that orbital moments are antiparallel to the spin moments, even though the $3d$ band is more than half-filled, a filling for which the orbital moment is expected to align parallel to the spin.
This anomaly may suggest that ferric-coupled Fe atoms exhibit an enhanced relative orbital contribution, analogous to the behavior previously reported for magnetite \cite{Goering2011-fo}.

\section*{Conclusion}
We have implemented the on‑the‑fly XMCD protocol on a motorized stage to map element‑resolved spin and orbital moments across a Fe–Ni–Co composition‑gradient combinatorial film. 
Despite the high scan speed, spectra processed with the Savitzky–Golay filter retained sufficient quality for quantitative magnetic analysis, enabling a comprehensive ternary map of magnetic trends.
Since the primary goal of this work is methodological, namely, the development of a high‑throughput measurement platform, more exhaustive materials characterization is still required to evaluate magnetic properties fully. 
Bulk‑sensitive fluorescence‑yield XAS, magneto‑optic Kerr microscopy for moment imaging, and X‑ray diffraction for crystal‑structure identification are obvious next steps. 
By integrating such complementary techniques into a unified data‑analysis pipeline, we envisage a rapid‑screening materials ecosystem that will accelerate magnetic‑materials discovery far beyond the capabilities of conventional, single‑point, and/or stepwise spectroscopy measurements.

\begin{acknowledgments}
We thank M. Takata and T. Arima for valuable discussions that contributed to this work.
This project is supported in part by the MEXT Program: Data Creation and Utilization-Type Material Research and Development Project (Digital Transformation Initiative Center for Magnetic Materials; JPMXP1122715503) and by the Japan Society for the Promotion of Science (JSPS) KAKENHI (JP19H04399, JP23K17145, JP24K03205, JP24H01685, JP24K17603, and JP25K0338).
This work was supported by MEXT Quantum Leap Flagship Program (MEXT Q-LEAP) Grant Number JPMXS0118068681.
This work is also partially supported by PRESTO(JPMJPR2102) and CREST(JPMJCR1861 and JPMJCR2435), Japan Science and Technology Agency (JST).
\end{acknowledgments}

\bibliography{STAM-M.bbl}

\end{document}